\documentclass[journal,twoside,web]{ieeecolor}
\usepackage{generic}
\usepackage{cite}
\usepackage{amsmath,amssymb,amsfonts}
\usepackage{algorithmic}
\usepackage{graphicx}
\usepackage{algorithm,algorithmic}
\usepackage{hyperref}
\hypersetup{hidelinks=true}
\usepackage{textcomp}
\usepackage{multirow}
\usepackage{subcaption}
\usepackage{pifont}
\usepackage{xcolor}
\usepackage{arydshln}

\def\BibTeX{{\rm B\kern-.05em{\sc i\kern-.025em b}\kern-.08em
    T\kern-.1667em\lower.7ex\hbox{E}\kern-.125emX}}

\begin{document}

\title{Privacy-Preserving Collaborative Medical Image Segmentation Using Latent Transform Networks}

\author{Saheed Ademola Bello, Muhammad Shahid Jabbar, Muhammad Sohail Ibrahim, Shujaat Khan 
\thanks{S.A. Bello and S. Khan are with the Department of Computer Engineering, College of Computing and Mathematics, King Fahd University of Petroleum \& Minerals, Dhahran 31261, Saudi Arabia. \\ M.S. Jabbar and S. Khan are with the SDAIA-KFUPM Joint Research Center for Artificial Intelligence, King Fahd University of Petroleum \& Minerals, Dhahran 31261, Saudi Arabia. \\ M.S. Ibrahim is with Interdisciplinary Research Center for Intelligent Secure Systems (IRC-ISS), King Fahd University of Petroleum \& Minerals, Dhahran 31261, Saudi Arabia. e-mail: shujaat.khan@kfupm.edu.sa). }}

\maketitle

\begin{abstract}
Collaborative training across multiple institutions is becoming essential for building reliable medical image segmentation models. However, privacy regulations, data silos, and uneven data availability prevent hospitals from sharing raw scans or annotations, limiting the ability to train generalizable models. Latent-space collaboration frameworks such as privacy-segmentation framework (SF) offer a promising alternative, but such methods still face challenges in segmentation accuracy and vulnerability to latent inversion and membership-inference attacks. This work introduces a privacy-preserving collaborative medical image segmentation framework (PPCMI-SF) designed for heterogeneous medical datasets. The approach combines skip-connected autoencoders for images and masks with a keyed latent transform that applies client-specific orthogonal mixing and permutation to protect latent features before they are shared. A unified mapping network on the server-side performs multi-scale latent-to-latent translation, enabling segmentation inference without exposing raw data. Experiments on four datasets: PSFH ultrasound, ultrasound nerve segmentation, FUMPE CTA, and cardiac MRI show that the proposed PPCMI-SF consistently achieves high Dice scores and improved boundary accuracy, as reflected by lower 95th percentile Hausdorff distance (HD95) and average symmetric surface distance (ASD) compared to the current state-of-the-art and performs competitively with privacy-agnostic baselines. Privacy tests confirm strong resistance to inversion and membership attacks, and the overall system achieves real-time inference with low communication overhead. These results demonstrate that accurate and efficient medical image segmentation can be achieved without compromising data privacy in multi-institution settings.
\end{abstract}

\begin{IEEEkeywords}
Privacy-preserving medical image segmentation, Federated learning, Keyed Latent Transform (KLT), Autoencoder-based feature encoding, Membership inference attack.
\end{IEEEkeywords}

\section{Introduction}
\IEEEPARstart{D}{eep} learning has become central to modern medical imaging, especially for segmentation tasks that support diagnosis, surgical planning, and treatment monitoring \cite{chen2024privacy}. Convolutional neural network (CNN) and transformer-based models achieve state-of-the-art performance, however their performance depends on access to large and diverse annotated datasets. In practice, such datasets are difficult to obtain. Strict privacy regulations, institutional data silos, and uneven data availability limit the sharing of raw medical images and annotations across hospitals. These restrictions increase the risk of biased models that generalize poorly across populations, scanners, and clinical environments \cite{zhu2024privacy}. As a result, there is a growing need for segmentation methods that preserve privacy while enabling collaborative learning across multiple institutions.

Several research directions have attempted to address this challenge. Federated learning (FL) \cite{yahiaoui2024federated, gupta2023collaborative, kanhere2024privacy, haripriya2025privacy, skorupko2025federated} allows models to be trained without sharing raw data, but it remains vulnerable to gradient inversion and membership inference attacks that can reveal patient-specific details. FL also relies on frequent communication rounds, which can introduce latency and make large-scale clinical deployment difficult. Encryption-based methods such as homomorphic encryption and secure multi-party computation \cite{ziller2020privacy, kiya2022privacy} offer stronger theoretical guarantees, yet they require large computational and memory resources, limiting their use in real-time medical imaging. Trusted execution environments \cite{bian2021privacy} reduce some of this overhead but depend on specialized hardware and are susceptible to side-channel attacks.

A practical alternative is to shift collaboration into the latent space. Encoding-based frameworks aim to separate identity-related information from task-relevant features. Privacy-segmentation framework (Privacy-SF) \cite{chen2024privacy} follows this idea by encoding both images and masks into latent representations on the client side, sending only these latents to a central server for learning. This reduces communication cost and avoids direct exposure of raw data. However, this work still faces two key limitations. First, segmentation fidelity is limited by low-resolution bottleneck latent representations, which can suppress fine spatial detail and boundary information during encoding. This can be improved by enhancing spatial feature recovery within the encoder–decoder structure. Second, the latent representations may still be vulnerable to inversion attacks, where an adversary attempts to reconstruct the original image using mismatched or auxiliary decoders.

\begin{figure*}[!ht]
    \centering
    \includegraphics[width=0.95\linewidth]{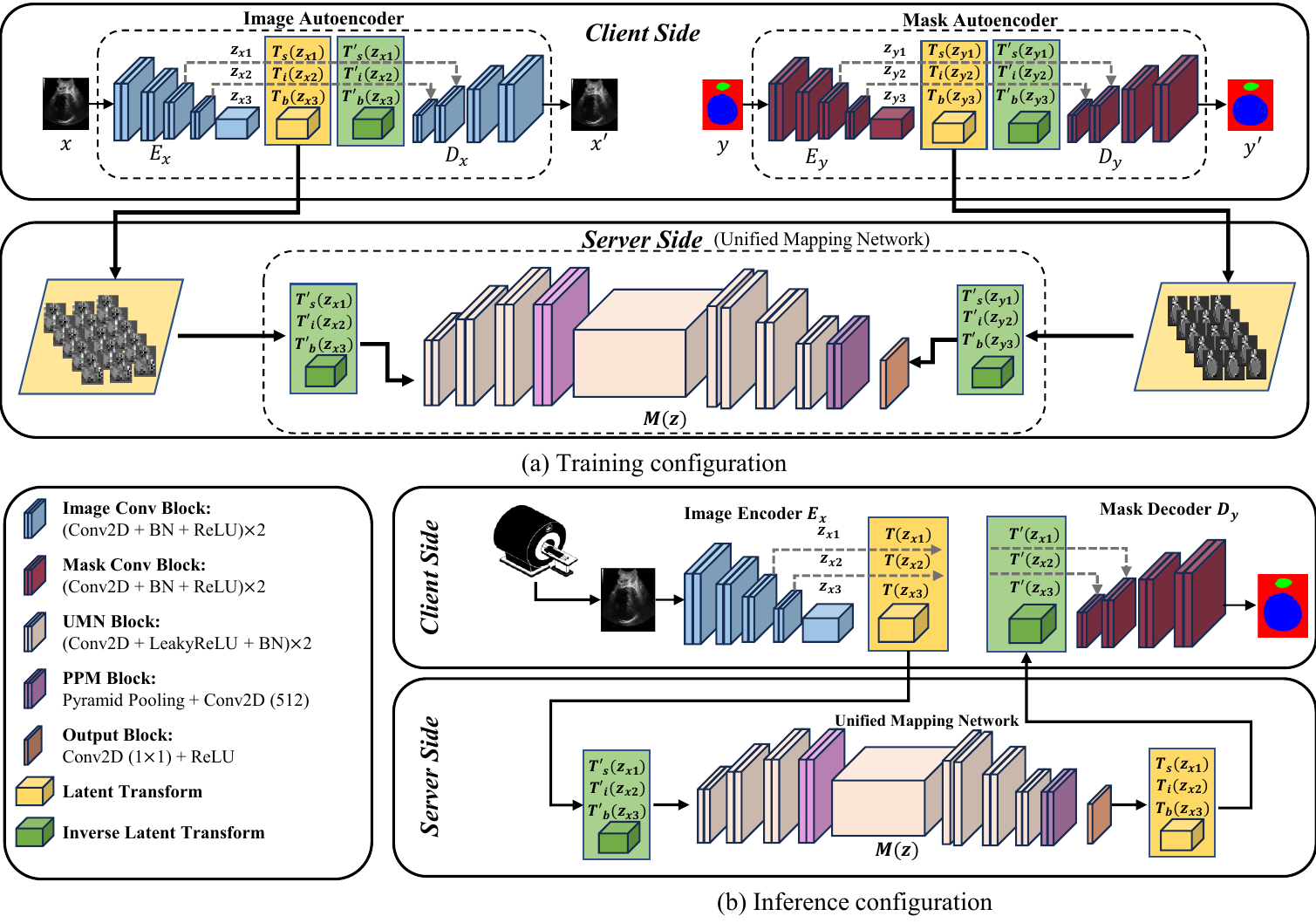}
    \caption{
    Overview of the proposed privacy-preserving collaborative medical image segmentation framework (PPCMI-SF). 
    \textbf{(a) Training configuration:} Each client trains its own image autoencoder $E_x$–$D_x$ and mask autoencoder $E_y$–$D_y$. 
    The resulting multi-scale latents $\{z_{x1},z_{x2},z_{x3}\}$ and $\{z_{y1},z_{y2},z_{y3}\}$ are secured using the client-keyed transforms 
    $\{T_s(\cdot), T_i(\cdot), T_b(\cdot)\}$ (yellow) before transmission.  
    Upon receiving the protected latents, the server applies the corresponding inverse transforms $\{T_s^{-1},T_i^{-1},T_b^{-1}\}$ (green) to recover the shared-domain latents and trains the unified mapping network $M(\cdot)$ to learn a latent-to-latent translation between \emph{unprotected} image and mask latents. 
    \textbf{(b) Inference configuration:} The client encodes an input image and applies the forward transforms $\{T_s,T_i,T_b\}$ to protect the latents before sending them to the server. The server again applies the private inverse transforms to map the recovered shared-domain latents through $M(\cdot)$, then re-applies the forward transforms to re-protect the predicted mask latents before returning them to the client. Finally, the client removes the protection, decodes the latents using $D_y$, and obtains the final segmentation mask.
    }

    \label{fig:proposed_model}
\end{figure*}

To address these limitations, a privacy-preserving collaborative medical image segmentation framework (PPCMI-SF), as illustrated in Fig.~\ref{fig:proposed_model}, is proposed. PPCMI-SF enhances both segmentation performance and latent-space privacy by incorporating two principal design strategies. Autoencoders equipped with skip-connections enhance feature propagation and preserve fine spatial structure without introducing significant computational overhead. In addition, a keyed latent transform (KLT) applies a lightweight, client-specific orthogonal mixing and permutation to protect latent features prior to transmission to the server. This transform makes the latents difficult to invert while remaining compatible with end-to-end learning. Together, these components improve the balance between privacy and utility by preserving both region overlap and boundary fidelity in the segmentation outputs. Skip-connections help maintain structural fidelity in segmentation outputs, while the KLT protects encoded features against inversion and reconstruction attempts. Unlike encryption-based or federated approaches, the proposed framework remains computationally efficient and suitable for multi-center settings.

The key contributions of this work are as follows:
\begin{enumerate}
    \item A proposed PPCMI-SF combining skip-connected autoencoders with a keyed latent transform for secure latent-space collaboration.
    \item A unified mapping network with pyramid pooling and an inverted encoder–decoder hierarchy for improved latent-to-latent segmentation under privacy constraints.
    \item Comprehensive evaluation of privacy robustness through cross-decoder inversion and membership-inference experiments.
    \item Demonstration of generalization on diverse modalities including ultrasound, CT (FUMPE), and cardiac MRI, approaching privacy-agnostic baselines without exposing raw data.
    \item Analysis of computational and communication cost, along with implementation details for reproducibility.
\end{enumerate}
The remainder of the paper is organized as follows: Sec.~\ref{sec:lit_review} discusses the existing works related to privacy preserving segmentation task. Sec.~\ref{sec:methodology} details the proposed work, followed by Sec. \ref{sec:results} providing a detailed discussion on the results and finally, Sec. \ref{sec:conclusion} concludes the paper.


\section{Related Works} \label{sec:lit_review}

\begin{table*}[h]
\centering
\caption{Taxonomy of privacy-preserving methods for medical image segmentation and representative works across major methodological categories.}
\label{tab:taxonomy}
\renewcommand{\arraystretch}{1.3}
\resizebox{\textwidth}{!}{
\begin{tabular}{p{3.3cm}|p{5.2cm}|p{4.5cm}|p{4.1cm}}
\hline \hline
\textbf{Category} & \textbf{Representative Methods} & \textbf{Strengths} & \textbf{Limitations} \\
\hline \hline

Federated and Blockchain-Based Learning 
& FL \cite{gupta2023collaborative, skorupko2025federated, yahiaoui2024federated}, SegViz \cite{kanhere2024privacy}, PriMIA \cite{ziller2020privacy}, Blockchain-FL \cite{kumar2024privacy}, decentralized FL \cite{fang2024decentralised}
& Raw data remains local; improves cross-site generalization; blockchain improves auditability
& Vulnerable to gradient leakage; communication overhead; non-IID data instability \\ \hline

Encryption and Differential Privacy 
& HE/SMPC \cite{ziller2020privacy, kim2021privacy}, encrypted ViT/patch models \cite{kiya2022privacy}, Mixup-Privacy \cite{kim2023mixup}, Privacy-Net \cite{kim2020privacy}, DP-based learning \cite{ziller2021differentially, hasan2025privacy, pan2024feddp}
& Strong theoretical privacy guarantees; encrypted inference; DP prevents memorization of patient features
& High computation cost (HE/SMPC); DP noise reduces accuracy; requires specialized architectures (e.g., ViT) \\ \hline

Synthetic and Continual Learning 
& Synthetic continual segmentation \cite{xu2024privacy}, Mixup-based synthesis \cite{zhu2024mp}, fuzzy steganographic learning \cite{chen2025fused}
& Avoids direct data sharing; mitigates catastrophic forgetting; flexible for incremental updates
& Synthetic data may distort anatomy; limited accuracy for complex imaging modalities \\ \hline

Trusted Hardware and Secure Inference 
& MP-Net \cite{zhu2024mp}, Hybrid Secure Inference (HSI) \cite{bian2021privacy}
& Real-time encrypted inference; efficient secure computation using TEE
& Hardware dependence; vulnerability to side-channel attacks; high deployment cost \\ \hline

Encoding- and Mapping-Based Frameworks 
& Privacy-SF \cite{chen2024privacy}, latent-space anonymization, encoder–decoder masking strategies
& Low communication cost; practical; avoids raw data exposure; efficient for multi-client setups
& Limited segmentation fidelity; vulnerable to latent inversion; reduced spatial detail \\

\hline \hline
\end{tabular}
}
\end{table*}

Privacy-preserving medical image segmentation has become increasingly important as modern AI systems rely on sensitive multi-institutional datasets. To balance diagnostic accuracy with patient confidentiality, diverse strategies have been developed, including FL, blockchain-enhanced collaboration, encryption and differential privacy, synthetic and continual learning, trusted hardware, and encoding-based frameworks. A brief summary of the related works is presented in Table~\ref{tab:taxonomy}.

Federated learning enables collaborative training without sharing raw data \cite{yahiaoui2024federated, fang2024decentralised, pietrantoni2023segloc}, and has demonstrated improved cross-site generalization \cite{gupta2023collaborative, skorupko2025federated}. Blockchain integration further enhances auditability and tamper resistance in distributed aggregation \cite{kumar2024privacy}, while adaptive aggregation and selective parameter synchronization address data heterogeneity and partial annotations \cite{haripriya2025privacy, kanhere2024privacy}. However, FL remains vulnerable to gradient leakage, membership inference, reconstruction attacks, and often incurs substantial communication and synchronization overhead.

Encryption-based approaches protect data at the algorithmic level. Homomorphic encryption offers strong theoretical guarantees but introduces significant computational cost \cite{kim2021privacy}. Learnable encryption and transformer-based secure segmentation have also been explored to preserve privacy during inference \cite{kiya2022privacy}. Complementary lightweight transformations, such as Mixup-privacy and adversarial anonymization, aim to obscure sensitive content while maintaining segmentation quality \cite{kim2023mixup, kim2020privacy}. Differential privacy further mitigates patient-level information leakage by injecting calibrated noise into data or gradients \cite{ziller2021differentially, hasan2025privacy, pan2024feddp}, however, it requires careful tuning to achieve a balanced trade-off between privacy and utility. Alternative directions reduce reliance on direct patient data through synthetic and continual learning strategies \cite{xu2024privacy}, as well as embedding and steganographic-style transformations that generate privacy-preserving intermediate representations \cite{zhu2024mp, chen2025fused}. Trusted hardware solutions, including secure enclave–based inference frameworks \cite{zhu2024mp, bian2021privacy}, accelerate secure computation but introduce hardware dependencies and vulnerabilities to potential side-channel attacks.

More recently, encoding-based frameworks have emerged as a practical middle ground between cryptographic protection and gradient-sharing schemes. In this context, privacy-SF \cite{chen2024privacy} encodes images and masks into compact latent spaces before centralized mapping, reducing communication cost and direct exposure of visual content. 
However, such approaches may sacrifice spatial detail in the latent space, limiting segmentation performance, and remain susceptible to latent feature inversion if paired with an alternative or poorly isolated decoder, exposing potential security weaknesses.

Building upon this foundation, the proposed PPCMI-SF incorporates two key architectural enhancements. First, Autoencoders equipped with skip-connections improve spatial fidelity by promoting feature reuse between encoder and decoder stages, thereby strengthening segmentation accuracy without increasing computational complexity. Second, a keyed latent transform introduces client-specific permutation and orthogonal mixing of latent tensors prior to transmission, enhancing resistance to inversion while preserving differentiability for seamless optimization. Collectively, these components enable improved segmentation performance and stronger empirical privacy protection with minimal computational overhead.


\section{Materials and Methods}\label{sec:methodology}
\subsection{Datasets}
\label{sec:datasets}

To evaluate the performance and generalizability of the proposed PPCMI-SF, experiments were conducted on multiple public medical imaging datasets spanning diverse modalities, anatomical targets, and acquisition conditions. The primary dataset for training and optimization was the pubic symphysis and fetal head (PSFH) dataset \cite{chen2024psfhs}, while three additional datasets, ultrasound nerve segmentation \cite{montoya2016nerve}, FUMPE (pulmonary embolism CTA) \cite{masoudi2018fumpe}, and a cardiac MRI dataset \cite{antonelli2022medical}, were used to evaluate cross-modality generalization.

\paragraph{Pubic Symphysis and Fetal Head (PSFH) Dataset}
The PSFH dataset originates from the MICCAI 2023 pubic symphysis and fetal head segmentation and angle of progression challenge \cite{chen2024psfhs}. It comprises 4,700 expert-annotated grayscale ultrasound images, with 4,000 used for training and 700 for official testing. Pixel-level annotations delineate the fetal head and pubic symphysis, making it among the primary benchmarks for evaluating segmentation performance under privacy-preserving settings.

\paragraph{Ultrasound Nerve Segmentation Dataset}
The ultrasound nerve segmentation dataset (US Nerve Seg.) \cite{montoya2016nerve} contains 5,635 neck ultrasound images with binary nerve masks. Compared to PSFH, it exhibits stronger speckle noise and finer anatomical structures, providing a challenging benchmark for assessing robustness to small-object localization and texture variability.

\paragraph{Cardiac MRI Dataset}
The cardiac MRI dataset \cite{antonelli2022medical} includes short-axis MRI slices annotated for the left ventricle, right ventricle, and myocardium. Pronounced contrast variations and inter-patient variability make it an effective benchmark for assessing robustness under domain shifts and modality changes.

\paragraph{FUMPE Pulmonary Embolism CTA Dataset}
The FUMPE dataset \cite{masoudi2018fumpe} consists of 8,792 axial computed tomography angiography (CTA) slices from 35 patients, annotated by expert radiologists for pulmonary embolism regions. The presence of small and peripheral lesions, along with intensity inhomogeneity, makes it suitable for evaluating cross-modality generalization to volumetric CT data.

\subsection{Proposed Segmentation Framework}

An overview of PPCMI-SF is illustrated in Fig.~\ref{fig:proposed_model}. The architecture comprises client-side encoding modules and a server-side unified mapping network. On each client, two skip-connected autoencoders independently encode the medical image and its corresponding mask into multi-scale latent representations. The image encoder $E_x$ maps an input image $x \in \mathbb{R}^{C \times H \times W}$ (with $C=1$) to hierarchical latent representations $\{z_{x1}, z_{x2}, z_{x3}\}$, while the mask encoder $E_y$ transforms the ground-truth mask $y \in \mathbb{R}^{C \times H \times W}$ into the corresponding latent representations $\{z_{y1}, z_{y2}, z_{y3}\}$. These representations retain task-relevant semantics while suppressing identifiable information.

Prior to transmission, each latent tensor is processed by a client-specific keyed latent transform (KLT) $T(\cdot)$, which performs orthogonal mixing and bias-based permutation to enhance privacy. Only protected latents are transmitted to the server. For each client, the server applies the corresponding inverse transform $T^{-1}(\cdot)$ to recover the shared latent domain, performs image-to-mask mapping via the unified mapping network, and re-applies $T(\cdot)$ before returning predictions. This design enables collaborative learning without exposing raw data or unprotected latent features.

During inference, a new image is encoded and transformed as $T(E_x(x))$ before transmission. The server computes predictions in the shared space and returns protected mask latents, which are inverted and decoded locally to produce the final segmentation:
\[
\hat{y} = D_y\!\left( T^{-1}\!\left( T \circ M_{\text{shared}} \circ T^{-1}\!\bigl( T(E_x(x)) \bigr) \right) \right).
\]

Throughout both training and inference, raw images, masks, and unprotected latents remain confined to the client side. Only transformed latent features are exchanged, ensuring secure multi-institutional collaboration. A multi-threaded client implementation further improves computational efficiency and scalability across participating centers.

\subsection{Client-Side Autoencoder Design}
\label{autoencoder-design}

The image and mask encoding networks in the proposed framework are implemented as skip-connected autoencoders that extract and reconstruct hierarchical feature representations from medical images and their segmentation masks. Each network follows a modified UNet architecture, optimized for privacy-preserving representation learning through multi-scale latent encoding and spatially aware reconstruction. The full architecture of the image and mask encoding networks is summarized in Table \ref{tab:encoder_architecture}.

\begin{table}[th]
\centering
\caption{Architecture of the image and mask autoencoders (output channels = 3 for PSFH dataset).}
\label{tab:encoder_architecture}
\renewcommand{\arraystretch}{1.15}
\setlength{\tabcolsep}{3pt}
\begin{tabular}{p{1.2cm}|p{1.1cm}|p{3.4cm} @{\hspace{10pt}}|p{2.1cm}}
\hline \hline
\textbf{Part} & \textbf{Block} & \textbf{Operation} & \textbf{Output Shape} \\
\hline \hline
Encoder & Input 
& [Conv(3$\times$3, s=2) + BN + ReLU] $\times$2 
& (16, H, W) \\[3pt]

& Down $\times$4 
& Downsample + [Conv + BN + ReLU] $\times$2 
& (32, H/2, W/2) $\rightarrow$ (256, H/16, W/16) \\
\hline

Latent Transform & $T_{b}(\cdot)$, $T_{i}(\cdot)$, \& $T_{s}(\cdot)$ 
& Orthogonal transform + bias $(Q,b)$, client-specific, non-trainable 
& -- \\
\hline

Decoder & Up $\times$4 
& Upsample + [Conv + BN + ReLU] $\times$2 
& (128, H/8, W/8) $\rightarrow$ (16, H, W) \\[3pt]

& Output 
& Conv(1$\times$1) + Sigmoid 
& (3, H, W) \\
\hline \hline
\end{tabular}
\end{table}

\subsubsection*{Encoder}

The encoder $E_x$ transforms an input image $x \in \mathbb{R}^{C \times H \times W}$ into three latent feature maps $\{z_{x1}, z_{x2}, z_{x3}\}$, corresponding to bottleneck, intermediate, and shallow levels of abstraction. Each encoding block consists of a $3 \times 3$ convolutional layer with stride 2, followed by batch normalization (BN) and a ReLU activation. Progressive downsampling in the encoder network extracts both semantic and structural cues while reducing redundant or irrelevant information.

\subsubsection*{Latent transformation}
After encoding, each latent representation passes through the client-specific keyed transform $T_{\alpha}(\cdot)$, explained in Sec.~\ref{subsec:latent_transform}, which performs orthogonal channel mixing and bias-based permutation to secure the latent features before transmission. The transformed tensors $\{T_b(z_{x1}), T_i(z_{x2}), T_s(z_{x3})\}$ are then transmitted to the server-side unified mapping network. Here, $T_{b}(\cdot)$, $T_{i}(\cdot)$ and $T_{s}(\cdot)$ represent the transformation functions for the bottleneck, intermediate, and shallow layers, respectively. Each KLT function corresponds to a distinct instantiation of the keyed transform tailored to the multi-scale latent structure of the architecture.

\subsubsection*{Decoder}

The decoder $D_x$ reconstructs the encoded input by sequentially upsampling feature maps using bilinear interpolation (scale factor 2), followed by $3 \times 3$ Conv-BN-ReLU layers. Skip connections link corresponding encoder and decoder layers to enhance spatial consistency and preserve fine structural details. The final layer applies a $3 \times 3$ convolution with a sigmoid activation, producing a reconstructed output $\hat{x}$ normalized to $[0, 1]$.

\subsubsection*{Mask autoencoder network}

The mask autoencoder network $(E_y, D_y)$ adopts the same architectural design as the image autoencoder but operates on segmentation masks rather than intensity images. The final output layer generates three channels $(N = 3)$ corresponding to the segmentation classes: background, fetal head, and pubic symphysis. After training, the encoded mask latent representations ${z_{y1}, z_{y2}, z_{y3}}$ are used as supervisory targets for the unified mapping network on the server side.

\subsection{Keyed Latent Transform}
\label{subsec:latent_transform}

To ensure secure communication of feature representations between clients and the server, a keyed latent transform (KLT) layer is embedded within the encoding pipeline. The KLT perturbs latent representations in a controlled and reversible manner and is applied after each encoder stage, ensuring that all transmitted features remain in protected latent form.

\begin{figure}[t]
    \centering
    \begin{subfigure}[t]{0.24\textwidth}
        \centering
        \includegraphics[width=\textwidth]{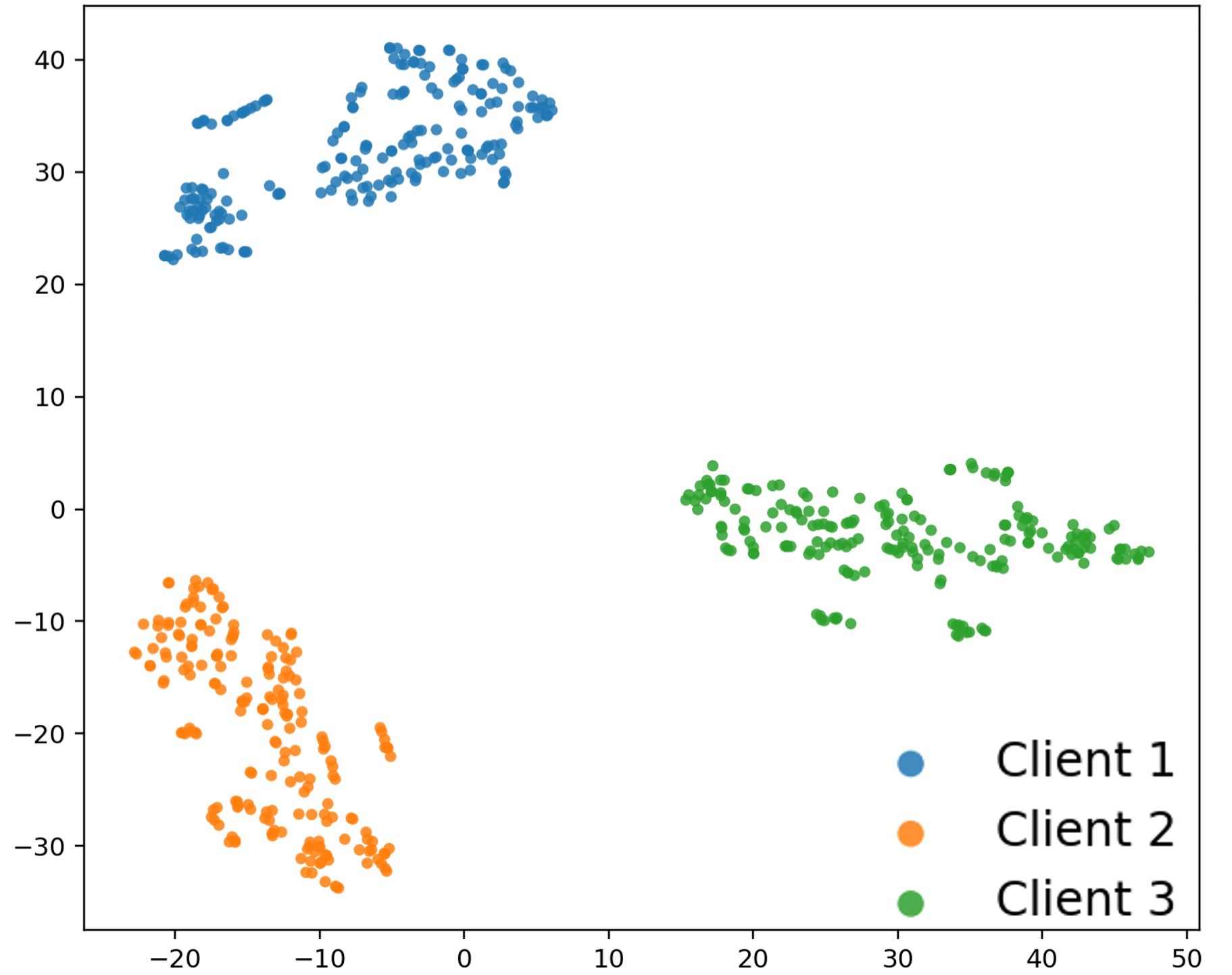}
        \caption{t-SNE before KLT}
        \label{fig:tsne_before}
    \end{subfigure}
    \hfill
    \begin{subfigure}[t]{0.24\textwidth}
        \centering
        \includegraphics[width=\textwidth]{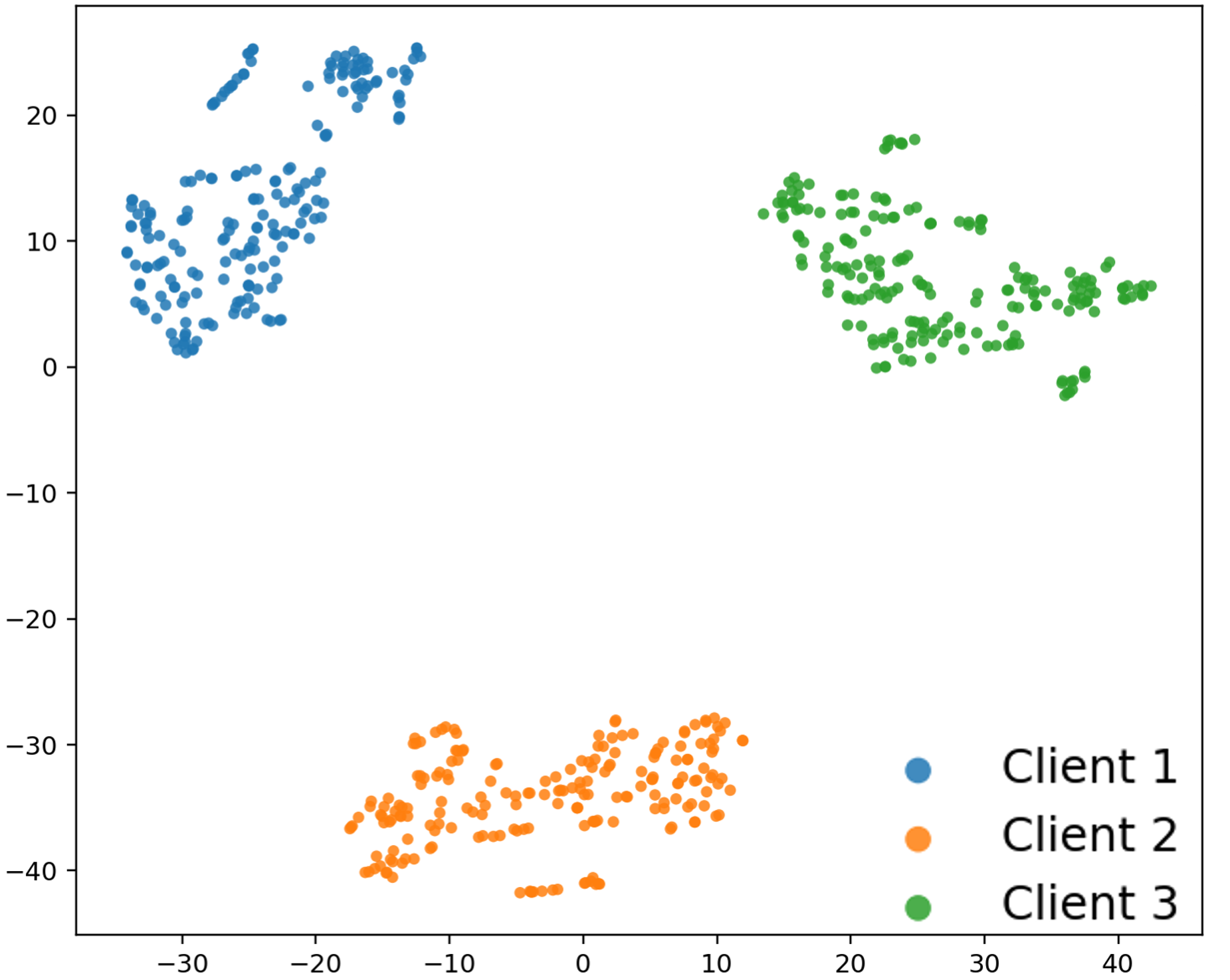}
        \caption{t-SNE after KLT}
        \label{fig:tsne_after}
    \end{subfigure}

    \caption{Visualization of latent distributions before and after the keyed latent transform (KLT).\\
    }
        \label{fig:klt_tsne}
\end{figure}

Formally, for a latent tensor \(z \in \mathbb{R}^{B \times C \times H \times W}\), the KLT is defined as
\[
z' = T(z) = Q^{\top} z + b,
\]
where \(Q \in \mathbb{R}^{C \times C}\) is a client-specific orthogonal matrix generated using QR decomposition \cite{gander1980algorithms}, and \(b \in \mathbb{R}^{C}\) is a bias vector broadcast across spatial dimensions. The corresponding inverse KLT is defined as
\[
z = T^{-1}(z') = Q \,(z' - b).
\]
Each client is associated with a unique key pair \((Q,b)\), while the server maintains the corresponding transforms to map between protected and shared latent domains. Unlike cryptographic encryption, the KLT preserves linearity and differentiability, enabling independent training of the server-side mapping network without cross-institutional end-to-end optimization.

As illustrated in Fig.~\ref{fig:tsne_before}, latent features exhibit client-specific geometric signatures prior to transformation. After applying KLT, the feature distributions become more uniform and geometrically consistent, reducing identifiable structural patterns while preserving semantic relationships as shown in Fig.~\ref{fig:tsne_after}. This orthogonal mixing and controlled translation mitigate client profiling risks without degrading downstream mapping performance. The KLT is computationally lightweight, involving only matrix multiplication and bias addition, thereby introducing negligible latency during training and inference. Compared to homomorphic encryption and related cryptographic schemes, it provides practical feature-level security while maintaining real-time efficiency and segmentation accuracy in multi-client medical settings.

\subsection{Unified Mapping Network (UMN) Architecture}
\label{sec:umn_architecture}

Conventional medical image segmentation networks are often based on encoder-decoder architectures in which features are progressively downsampled to capture high-level semantics and then upsampled to recover spatial resolution. In the proposed PPCMI-SF, the inputs to the unified mapping network (UMN) are already encoded representations produced by the client-side autoencoders, therefore, further downsampling can be redundant and may degrade spatial precision and localization accuracy. To address this, we incorporate an overcomplete UMN architecture where intermediate feature maps are expanded in spatial resolution relative to the input and output, allowing finer contextual reasoning within the latent space.

The proposed UMN inverts the roles of a conventional encoder-decoder pair. Its encoder branch employs bilinear interpolation for progressive upsampling, whereas the decoder branch integrates max-pooling layers for controlled spatial reduction. This inversion helps preserve spatial detail along the transformation path and reduces the information loss typical of aggressive downsampling. To capture global contextual dependencies, the proposed UMN incorporates a pyramid pooling module (PPM) \cite{ zhao2017pyramid} at several stages
The PPM aggregates context using four pooling scales, $1\times1$, $3\times3$, $5\times5$, and $7\times7$, which are then upsampled and concatenated with the base feature map. This multi-scale aggregation improves semantic representation and facilitates alignment between global and local structures in the protected latent domain. Skip connections between encoder and decoder branches further support recovery of boundary-related and region-specific cues.

Formally, the UMN $M(\cdot)$ receives protected image latent features and produces protected mask latent features,
\[
\hat{T}(z_y) = T \circ M_{\text{shared}} \circ T^{-1} \circ T(z_x).
\]

where the input and output features share identical dimensions $z_x, z_y \in \mathbb{R}^{c\times h\times w}$. The dimensions for bottleneck, intermediate, and shallow layers are $256 \times 8 \times 8$, $128 \times 16 \times 16$, and $64 \times 32 \times 32$ respectively. The layer configuration of the proposed UMN is summarized in Table \ref{tab:mapping_network_architecture}.

\begin{table}[th]
\centering
\caption{Layer-wise configuration of the proposed unified mapping network (UMN).}
\label{tab:mapping_network_architecture}
\renewcommand{\arraystretch}{1.3}
\setlength{\tabcolsep}{3pt}
\begin{tabular}{p{1.2cm}| p{1.6cm}| p{5cm}}
\hline \hline
\textbf{Stage} & \textbf{Block} & \textbf{Operation and Output} \\
\hline \hline

Input & Input tensor &
Encoded latent $(h,w,c)$;\newline e.g., $(8,8,256)$ for bottleneck input
\\ \hline

Encoder 
& Conv $\times$3 
& Conv(3$\times$3): 128$\rightarrow$256$\rightarrow$512,
\newline LeakyReLU, BN; Dropout(0.2) after 512 \\ \hline

& PPM 
& Pyramid pooling (1,3,5,7) + Conv(512) \\ \hline

Decoder 
& Conv $\times$2 
& Conv(3$\times$3): 512$\rightarrow$1024, 
LeakyReLU,\newline BN; Dropout(0.2) after 1024 \\ \hline

& Downsample $\times$3 
& MaxPool(2$\times$2) + Conv: 
256$\rightarrow$128$\rightarrow$64;\newline
output scales $(h/2,w/2)\rightarrow(h/8,w/8)$ \\ \hline

Upsample 
& Up $\times$3 
& Upsample(2$\times$) + skip +\newline Conv: 
128$\rightarrow$256$\rightarrow$512;\newline
PPM(512) in final block \\ \hline

Output 
& Output layer 
& Conv(1$\times$1) + ReLU $\rightarrow$ $(h,w,c)$ \\

\hline \hline
\end{tabular}
\end{table}

The design enables the UMN to preserve rich spatial resolution throughout the transformation pathway while capturing both fine-grained and coarse-grained semantic correspondences. The PPM improves contextual awareness, and the reversed encoder–decoder hierarchy mitigates information loss compared to heavily downsampling architectures. By operating directly on protected latent representations rather than raw pixels, the network attains high segmentation fidelity while maintaining strong privacy guarantees.


\subsection{Training Strategy, Multi-Client Collaboration, and Loss Functions}
\label{sec:training_strategy}

The proposed framework follows a two-stage client-server training paradigm. Image and mask autoencoders are trained locally at each participating institution/client, while the UMN is trained centrally on the server using only latent representations. The two-stage training strategy enables distributed learning across multiple medical centers without exchanging raw data, ensuring confidentiality and privacy during training and inference.

\paragraph{Client-side training}
Each client independently trains two autoencoders: the image autoencoder $(E_x, D_x)$ and the mask autoencoder $(E_y, D_y)$. These networks are optimized to learn semantically rich yet privacy-preserving latent representations that capture both global context and local structural details relevant to segmentation. To balance pixel-level reconstruction accuracy with region-level structural consistency, a loss fusion of binary cross-entropy (BCE) and Dice loss is adopted:
\[
\mathcal{L}_{AE} = \alpha \cdot \text{BCE}(u, \hat{u}) + (1-\alpha) \cdot \text{Dice}(u, \hat{u}),
\]
where $u$ and $\hat{u}$ denote the ground-truth and reconstructed outputs, respectively, and $\alpha$ denote the loss fusion parameter. This formulation promotes stable structural recovery while limiting overfitting to fine-grained intensity patterns that may compromise privacy.

After training, the encoders $E_x$ and $E_y$ generate three levels of latent representations, ${z_{x1}, z_{x2}, z_{x3}}$ and ${z_{y1}, z_{y2}, z_{y3}}$, respectively. Each latent tensor is subsequently processed by the KLT $T(\cdot)$ prior to transmission, ensuring privacy in the latent domain.


\paragraph{Server-side training}
After receiving the protected latent pairs $[{T(z_{x\ell}), T(z_{y\ell})}]$ from all clients, the server applies the corresponding client-specific inverse transform to recover the shared latent domain:
\[
z_{x\ell} = T^{-1}(T(z_{x\ell})), \qquad 
z_{y\ell} = T^{-1}(T(z_{y\ell})).
\]
These inverted visually unrecognizable feature tensors are then used to train the UMN $M(\cdot)$, which learns a latent-to-latent translation between image and mask representations. The UMN is optimized using a multi-scale mean squared error,
\[
\mathcal{L}_{map} = \sum_{\ell \in \{1,2,3\}} 
\bigl\| z_{y\ell} - \hat{z}_{y\ell} \bigr\|_2^2,
\]
where $\hat{z}_{y\ell}$ are the predicted mask latents. After the forward pass, the server re-applies the KLT to produce protected outputs,
\[
\hat{T}(z_{y\ell}) = T(\hat{z}_{y\ell}),
\]
which are the only tensors transmitted back to the corresponding client. This design guarantees that unprotected latent representations remain strictly within the server environment. The UMN is trained using the Adam optimizer with a learning rate of $10^{-4}$ and early stopping with a patience of 20 epochs based on validation performance.

\begin{figure*}[ht!]
    \centering
    \includegraphics[width=\textwidth]{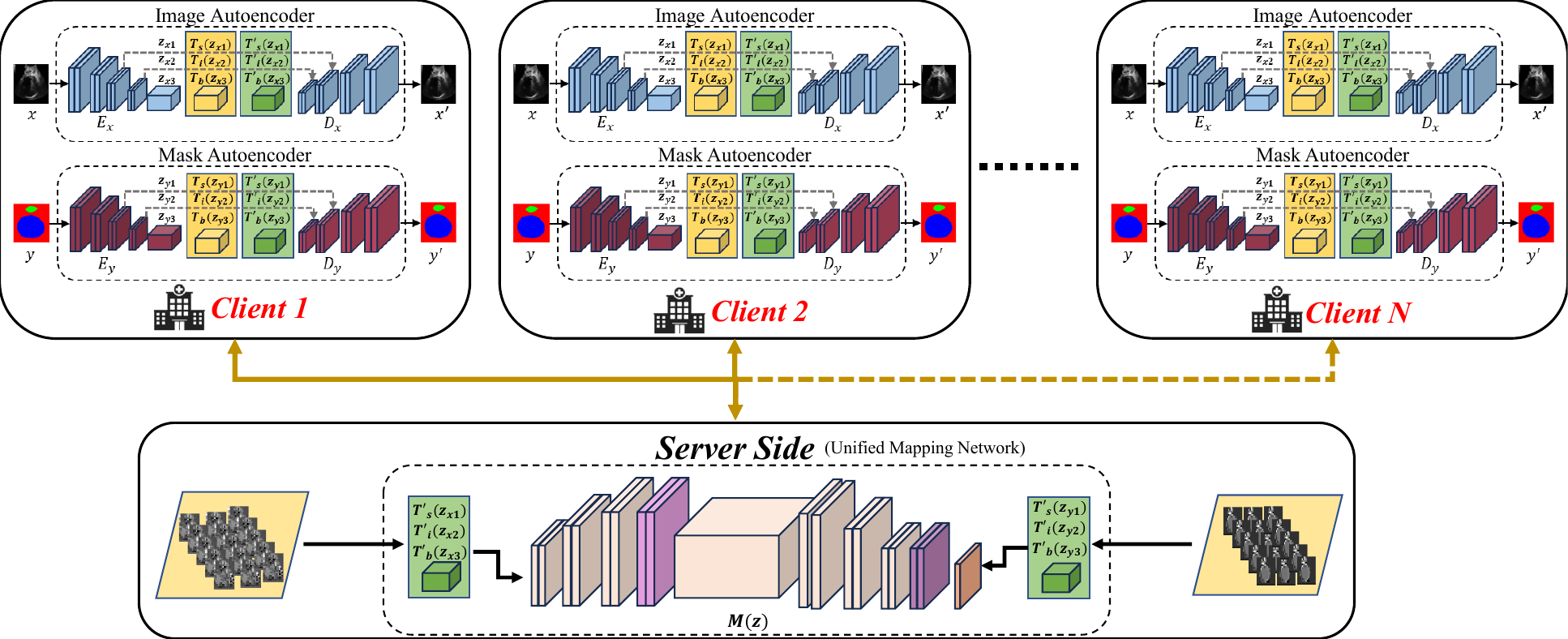}
    \caption{Multi-center collaboration framework of the proposed PPCMI-SF. Client institutions train local autoencoders and send only protected latents to the central server, which learns a unified mapping network for segmentation. 
    }
    \label{fig:gen_framework}
\end{figure*}

\paragraph{Multi-client collaboration}
The proposed PPCMI-SF supports scalable collaboration among multiple clients through parallelized local training and centralized model fusion as illustrated in Fig. \ref{fig:gen_framework}. Each client runs in an isolated training thread and optimizes its autoencoders on locally stored medical images and masks. Once training is complete, the client transmits only the KLT features and their associated mask KLT features to the central server. The server aggregates these transformed features into a global dataset used to train the UMN, effectively enabling a federated-style collaboration without the frequent communication and gradient exchange typical of conventional federated learning.

\paragraph{Privacy-aware training efficiency}
Thread-level parallelism at the client side reduces total training time while preserving data locality, as all image and mask processing is performed locally. Each client retains its own KLT parameters $(Q,b)$, which are not shared with other clients, while the server applies only the corresponding inverse KLT for training the shared UMN. The server does not access raw data, and the received KLT representations remain non-visual even after inversion, preventing exposure of identifiable patient information. The orthogonality of the transforms further ensures numerical stability and avoids distortion or amplification of latent statistics during optimization. Overall, the training strategy integrates distributed autoencoder learning with centralized latent-space mapping enabling promising segmentation performance, robust cross-domain generalization, and improved privacy protection in realistic multi-client environments.



\subsection{Threat Model and Privacy Objectives}
\label{sec:threatmodel}

\textbf{Adversary Model:}  
The threat model assumes an honest-but-curious central server and a potential external adversary. The server is authorized to execute the UMN and apply client-specific inverse KLT during training and inference, yet it is not granted access to raw medical images, segmentation masks, or unprotected latent representations. An external adversary may observe protected latent features, server-side model parameters, and prediction outputs, but does not have access to client-side data or the client-specific transform keys. Under this setting, potential attack vectors include (i) latent inversion using auxiliary decoders, (ii) cross-decoder reconstruction across clients, and (iii) membership-inference attacks targeting the server-side model.


\textbf{Assets and Assumptions:}  
Raw images, segmentation masks, and unprotected latent representations remain strictly on the client side at all times. Each client maintains private KLT parameters $(Q,b)$, which are not shared across clients. The server applies the corresponding inverse KLT only to recover a shared latent domain for mapping and does so within a protected execution context or equivalent trusted computing boundary. Key compromise, malicious server behavior, physical access, and side-channel attacks are considered out of scope for this study and may be considered for future work.

\textbf{Privacy Objectives:} The proposed PPCMI-SF is designed to:
\begin{enumerate}
    \item Prevent reconstruction of identifiable visual or anatomical content using the transmitted latent representations.
    \item Block meaningful decoding of latent features by unauthorized or cross-client decoders.
    \item Reduce membership leakage such that training samples cannot be reliably distinguished from non-members.
\end{enumerate}

\textbf{Privacy Mechanism and Limitations:}  
Privacy in PPCMI-SF is achieved through architectural isolation and latent-space obfuscation rather than formal cryptographic guarantees. The KLT applies client-specific orthogonal mixing that disrupts latent alignment across clients, thereby reducing susceptibility to inversion and membership-inference attacks while preserving segmentation utility. Accordingly, the privacy guarantees provided by PPCMI-SF are empirical and attack-specific, reflecting strong resistance under the evaluated threat model. Extending the framework to handle key compromise or fully malicious servers represents an important direction for future work.

\subsection{Experimental Setup}
\label{sec:inference_evaluation}

After training the client-side and server-side networks, the proposed framework performs inference in a fully privacy-preserving manner. The inference pipeline mirrors the communication structure used during training but removes all backward information flow, ensuring that no gradients or raw image data are exposed to the server. All encoding, decoding, and KLT steps remain strictly local to each client, while the server operates exclusively on protected latent representations.

\subsubsection{Inference procedure}
Given an unseen medical image $x$, the client first forwards it to the trained image encoder $E_x$ to extract multi-scale latent representations $\{z_{x1}, z_{x2}, z_{x3}\}$. These latent representations are then transformed using the client-specific KLT $T(\cdot)$, producing the protected features $\{T_b(z_{x1}), T_i(z_{x2}), T_s(z_{x3})\}$. Each protected feature corresponds to a different instantiation of the KLT tailored to the multi-scale latent structure of the encoder architecture. The client transmits these protected tensors to the server. 

The UMN $M(\cdot)$ on the server-side processes these inputs and predicts the corresponding protected mask latents $\{\hat{T}(z_{y1}), \hat{T}(z_{y2}), \hat{T}(z_{y3})\}$. The predictions are sent back to the client, which applies its corresponding private inverse KLT $T^{-1}(\cdot)$ to recover the unprotected latent domain:
\[
(\hat{z}_{y1}, \hat{z}_{y2}, \hat{z}_{y3}) 
= 
\bigl(
T^{-1}(\hat{T}(z_{y1})),\,
T^{-1}(\hat{T}(z_{y2})),\,
T^{-1}(\hat{T}(z_{y3}))
\bigr).
\]
The final segmentation mask is reconstructed locally using the client-specific mask decoder $D_y$:
\[
\hat{y} = D_y(\hat{z}_{y1}, \hat{z}_{y2}, \hat{z}_{y3}).
\]
This protocol ensures that raw images, masks, and unprotected latent representations never leave the client environment, while the server remains restricted to processing privacy-preserving representations.

All baseline models were re-trained under a unified experimental protocol, including identical input resolution, data splits, loss functions, and optimization settings, to ensure a fair comparison.

\subsubsection{Quantitative evaluation metrics}

Segmentation performance is evaluated using Dice similarity coefficient (DSC), 95th percentile Hausdorff Distance (HD95), and Average Symmetric Surface Distance (ASD), which jointly assess region overlap and boundary accuracy. The DSC measures volumetric overlap between the prediction $P$ and ground truth $G$:
\[
\text{DSC} = \frac{2|P \cap G|}{|P| + |G|},
\]
with higher values indicating better segmentation quality.

Boundary accuracy is quantified using HD95. Let $\partial P$ and $\partial G$ denote the predicted and ground-truth surfaces. The directed Hausdorff distance is calculated as
\[
d_H(\partial P, \partial G) = \max_{p \in \partial P} \min_{g \in \partial G} \| p - g \|_2 .
\]
HD95 corresponds to the 95th percentile of the symmetric point-to-surface distances, reducing sensitivity to extreme outliers. Lower values of HD95 indicate improved boundary localization.

The ASD measures the average symmetric boundary discrepancy calculated as
\[
\text{ASD} =
\frac{1}{|\partial P| + |\partial G|}
\left(
\begin{aligned}
&\sum_{p \in \partial P} \min_{g \in \partial G} \| p - g \|_2 \\
&\quad + \sum_{g \in \partial G} \min_{p \in \partial P} \| g - p \|_2
\end{aligned}
\right)
\]
where $\|\cdot\|_2$ denotes the Euclidean distance. Lower ASD values correspond to smoother boundary alignment.

Reconstruction fidelity of autoencoders and cross-latent mappings is evaluated using structural similarity index measure (SSIM) and the peak signal-to-noise ratio (PSNR) defined respectively as:
\[
\text{SSIM}(x,\hat{x}) =
\frac{(2\mu_x\mu_{\hat{x}} + C_1)(2\sigma_{x\hat{x}} + C_2)}
     {(\mu_x^2 + \mu_{\hat{x}}^2 + C_1)(\sigma_x^2 + \sigma_{\hat{x}}^2 + C_2)},
\]
\[
\text{PSNR} = 10 \log_{10}\!\left(\frac{1}{\mathrm{MSE}(x,\hat{x})}\right).
\]
Here, $\mu_x$ and $\sigma_x$ represent the mean and variance of the reference image, and $C_1 = 1 \times 10^{-4}$ and $C_2 = 9 \times 10^{-4}$ are stability constants.

For membership-inference evaluation, sensitivity and specificity are reported:
\[
\text{Sensitivity} = \frac{TP}{TP + FN}, \qquad
\text{Specificity} = \frac{TN}{TN + FP},
\]
where $TP$, $TN$, $FP$, and $FN$ denote the numbers of true positives, true negatives, false positives, and false negatives, respectively, as defined by the standard classification confusion matrix.

Together, Dice, HD95, and ASD provide a comprehensive assessment of segmentation performance by capturing both regional overlap and fine-grained boundary fidelity under privacy-preserving constraints. SSIM and PSNR complement this analysis by quantifying perceptual reconstruction quality of latent representations, while sensitivity and specificity measure resistance to membership-inference attacks.


\section{RESULTS AND DISCUSSION} \label{sec:results}
The primary evaluation of PPCMI-SF was conducted on the PSFH dataset. For consistency and fairness, the proposed PPCMI-SF and the baseline methods were trained under a unified training protocol, including identical image normalization procedures, a fixed input resolution of ($1 \times 256 \times 256$), and the same optimization objectives.

\subsection{Autoencoder Reconstruction Performance}
\label{sec:autoencoder_results}

The proposed PPCMI-SF relies on well-optimized local autoencoders to enable each client in learning expressive yet privacy-preserving latent representations. Prior to training the server-side UMN and client-side segmentation models, the reconstruction capability of the image and mask autoencoders is evaluated, as it reflects the quality and stability of the learned latent spaces.

Table~\ref{tab:autoencoder_performance} summarizes the reconstruction performance of Privacy-SF compared to the proposed PPCMI-SF across three clients using PSNR and SSIM. The proposed framework consistently improves image reconstruction quality, increasing PSNR from 24.79,dB to 26.16,dB and SSIM from 0.49 to 0.71, indicating stronger structural consistency in the learned latent representations. For mask reconstruction, SSIM remains high at 0.99 for both methods, while PSNR increases from 30.60 dB to 34.83 dB, demonstrating more precise recovery of segmentation boundaries. These improvements highlight the effectiveness of the proposed skip-connected autoencoder design in preserving multi-scale spatial information and stabilizing latent feature learning across clients.


\begin{table}[!t]
\centering
\caption{Reconstruction performance of client-specific image and mask autoencoders (AEs) in Privacy-SF and PPCMI-SF. Higher PSNR and SSIM indicate better reconstruction quality.}
\label{tab:autoencoder_performance}
\footnotesize
\setlength{\tabcolsep}{3pt}
\renewcommand{\arraystretch}{1.05}
\begin{tabular}{c|c|cc|cc|cc}
\hline \hline
\multirow{2}{*}{\textbf{Model}} & \multirow{2}{*}{\textbf{AE}} 
& \multicolumn{2}{c}{\textbf{C1}} 
& \multicolumn{2}{c}{\textbf{C2}} 
& \multicolumn{2}{c}{\textbf{C3}} \\ 
\cline{3-8}
& 
& \textbf{PSNR} & \textbf{SSIM} 
& \textbf{PSNR} & \textbf{SSIM} 
& \textbf{PSNR} & \textbf{SSIM} \\ 
\hline \hline
\multirow{2}{*}{Privacy-SF} 
& Image AE & 24.62 & 0.49 & 24.68 & 0.48 & 24.79 & 0.49 \\
& Mask AE  & 30.60 & 0.99 & 30.22 & 0.99 & 30.35 & 0.99 \\ 
\hline
\multirow{2}{*}{PPCMI-SF} 
& Image AE & 26.19 & 0.70 & 25.53 & 0.69 & 26.16 & 0.71 \\
& Mask AE  & 34.76 & 0.99 & 34.83 & 0.99 & 33.47 & 0.99 \\ 
\hline \hline
\multicolumn{8}{l}{\footnotesize CX: Client-X; AE: Autoencoder.}
\end{tabular}
\end{table}

\begin{figure}[!ht]
\centering
\includegraphics[width = \linewidth]{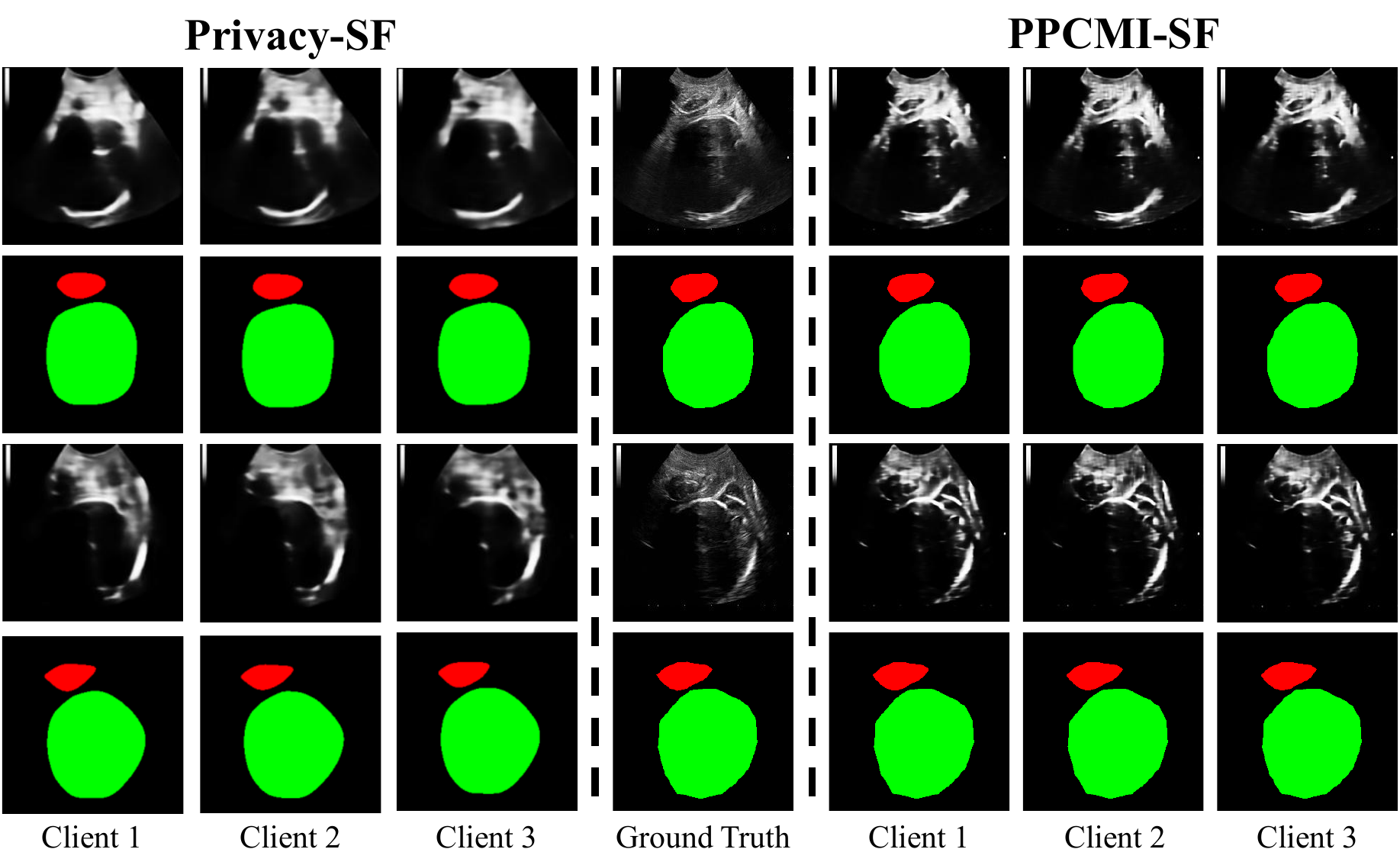} 
\caption{Qualitative reconstruction results from client-specific image and mask autoencoders. The proposed PPCMI-SF yields sharper and more structurally accurate reconstructions across all clients compared to the baseline Privacy-SF method.
}
\label{fig:autoencoder_visuals}
\end{figure}

Qualitative examples in Fig.~\ref{fig:autoencoder_visuals} further support the quantitative results presented in Table~\ref{tab:autoencoder_performance}. Image reconstructions generated by the proposed PPCMI-SF exhibit clearer tissue structures and improved contrast consistency, with visibly fewer artifacts compared to the Privacy-SF baseline. Similarly, the reconstructed masks present sharper object boundaries and more homogeneous regions, indicating stable structural recovery in the latent space. These observations suggest that the enhanced autoencoder design captures semantically meaningful features while maintaining robustness under privacy-preserving constraints, thereby providing a reliable foundation for subsequent latent-space mapping and segmentation.



\subsection{Performance on the PSFH Dataset}
\label{sec:psfh_results}

\renewcommand{\arraystretch}{1.2}
\begin{table*}[!ht]
\centering
\caption{Quantitative comparison of PPCMI-SF and state-of-the-art models on the PSFH dataset. Performance is reported using Dice Similarity Coefficient (DSC $\uparrow$), 95th percentile Hausdorff Distance (HD95 $\downarrow$), and Average Symmetric Surface Distance (ASD $\downarrow$).}
\label{tab:psfh_results}
\resizebox{\linewidth}{!}{
\begin{tabular}{c|ccc|ccc|ccc|c}
\hline \hline
\multirow{2}{*}{\bf Model} 
& \multicolumn{3}{c|}{\bf DSC $\uparrow$} 
& \multicolumn{3}{c|}{\bf HD95 $\downarrow$} 
& \multicolumn{3}{c|}{\bf ASD $\downarrow$} 
& \bf Param (M) \\ \cline{2-10}

& \bf FH & \bf PS & \bf PSFH 
& \bf FH & \bf PS & \bf PSFH 
& \bf FH & \bf PS & \bf PSFH 
&  \\ \hline \hline

UNet~\cite{ronneberger2015u} 
& $91.25 \pm 0.07$ & $82.80 \pm 0.17$ & $90.55 \pm 0.06$
& $17.25 \pm 0.18$ & $\mathbf{9.34 \pm 0.91}$ & $16.99 \pm 0.16$
& $5.55 \pm 0.77$ & $3.15 \pm 0.31$ & $4.86 \pm 0.47$
& 31.03 \\ \hline

AttUNet~\cite{oktay2018attention}
& $91.10\pm0.42$ & $79.25\pm2.35$ & $90.08\pm0.36$
& $17.49 \pm 0.55$ & $13.86 \pm 1.40$ & $20.52 \pm 0.67$
& $5.13 \pm 0.23$ & $4.93 \pm 0.90$ & $4.76 \pm 0.17$
& 35.6 \\ \hline

nnUNetv2~\cite{isensee2021nnu}
& $92.50 \pm 0.17$ & $82.30 \pm 0.73$ & $91.50 \pm 0.18$
& $13.71 \pm 0.33$ & $10.79 \pm 0.77$ & $\mathbf{15.48 \pm 0.25}$
& $4.29 \pm 0.11$ & $3.75 \pm 0.27$ & $\mathbf{3.00 \pm 0.09}$
& 92.5 \\ \hline

TransUNet~\cite{chen2021transunet}
& $\mathbf{93.02 \pm 0.05}$ & $81.66 \pm 1.25$ & $91.95 \pm 0.13$
& $\mathbf{13.25 \pm 0.44}$ & $11.36 \pm 0.56$ & $15.78 \pm 0.45$
& $\mathbf{3.96 \pm 0.05}$ & $3.59 \pm 0.18$ & $3.76 \pm 0.08$
& 105.3 \\ \hline

SwinUNet~\cite{cao2022swin}
& $92.06 \pm 0.10$ & $82.84 \pm 0.19$ & $91.15 \pm 0.09$
& $13.96 \pm 0.07$ & $11.06 \pm 0.15$ & $16.14 \pm 0.09$
& $4.38 \pm 0.05$ & $3.45 \pm 0.08$ & $4.05 \pm 0.04$
& 27.2 \\ \hline

DSEUNet~\cite{li2024dseunet}
& $92.44 \pm 0.34$ & $84.86 \pm 0.23$ & $91.75 \pm 0.31$
& $16.47 \pm 0.90$ & $12.67 \pm 1.83$ & $18.01 \pm 0.79$
& $4.47 \pm 0.22$ & $3.35 \pm 0.12$ & $3.98 \pm 0.17$
& 62.9 \\ \hline

UPerNet~\cite{wang2023upernet}
& $92.24 \pm 0.15$ & $80.67 \pm 0.97$ & $91.09 \pm 0.19$
& $14.37 \pm 0.26$ & $12.60 \pm 0.91$ & $17.12 \pm 0.50$
& $4.43 \pm 0.08$ & $3.80 \pm 0.29$ & $4.20 \pm 0.08$
& 31.3 \\ \hline

MissFormer~\cite{huang2022missformer}
& $91.58 \pm 0.95$ & $80.61 \pm 2.33$ & $90.54 \pm 01.01$
& $15.78 \pm 1.11$ & $13.65 \pm 0.83$ & $18.67 \pm 1.31$
& $4.82 \pm 0.62$ & $3.97 \pm 0.57$ & $4.45 \pm 0.50$
& 42.3 \\ \hline

H2Former~\cite{he2023h2former}
& $92.64 \pm 0.29$ & $\mathbf{85.32 \pm 0.45}$ & $91.94 \pm 0.26$
& $14.06 \pm 0.22$ & $11.07 \pm 0.64$ & $16.25 \pm 0.14$
& $4.20 \pm 0.15$ & $2.97 \pm 0.14$ & $3.78 \pm 0.11$
& 33.9 \\ \hline

FATNet~\cite{wu2022fat}
& $92.94 \pm 0.15$ & $80.78 \pm 1.45$ & $91.90 \pm 0.21$
& $14.13 \pm 0.24$ & $12.45 \pm 0.83$ & $16.71 \pm 0.41$
& $3.99 \pm 0.07$ & $4.58 \pm 0.57$ & $3.80 \pm 0.10$
& 28.8 \\ \hline

CE-Net~\cite{zhou2022contextual}
& $92.68 \pm 0.44$ & $84.90 \pm 0.25$ & $\mathbf{91.96 \pm 0.38}$
& $14.57 \pm 0.89$ & $12.55 \pm 0.89$ & $16.21 \pm 0.92$
& $4.16 \pm 0.27$ & $3.33 \pm 0.28$ & $3.77 \pm 0.19$
& 29.0 \\ \hline

MedSAM~\cite{ma2024segment}
& $90.59 \pm 0.40$ & $83.85 \pm 1.08$ & $89.95 \pm 0.40$
& $16.37 \pm 0.98$ & $9.52 \pm 0.67$ & $17.19 \pm 0.83$
& $5.26 \pm 0.26$ & $\mathbf{2.79 \pm 0.17}$ & $4.54 \pm 0.20$
& 93.7 \\ \hline

Privacy-SF\cite{chen2024privacy}
& $89.15 \pm 0.08$ & $74.55 \pm 0.16$ & $87.60 \pm 0.07$
& $17.08 \pm 0.13$ & $12.20 \pm 0.14$ & $17.15 \pm 0.12$
& $6.22 \pm 0.05$ & $12.2 \pm 0.05$ & $5.76 \pm 0.04$
& $\mathbf{20.55}$ \\ \hline

Proposed PPCMI-SF
& $91.78 \pm 0.05$ & $80.82 \pm 0.15$ & $90.49 \pm 0.05$
& $16.74 \pm 0.18$ & $13.20 \pm 0.27$ & $16.97 \pm 0.16$
& $5.39 \pm 0.04$ & $4.02 \pm 0.03$ & $4.90 \pm 0.03$
& $20.65$ \\ \hline \hline

\end{tabular}
}
\end{table*}


\begin{figure}[t!]
\centering
\includegraphics[width=0.75\linewidth]{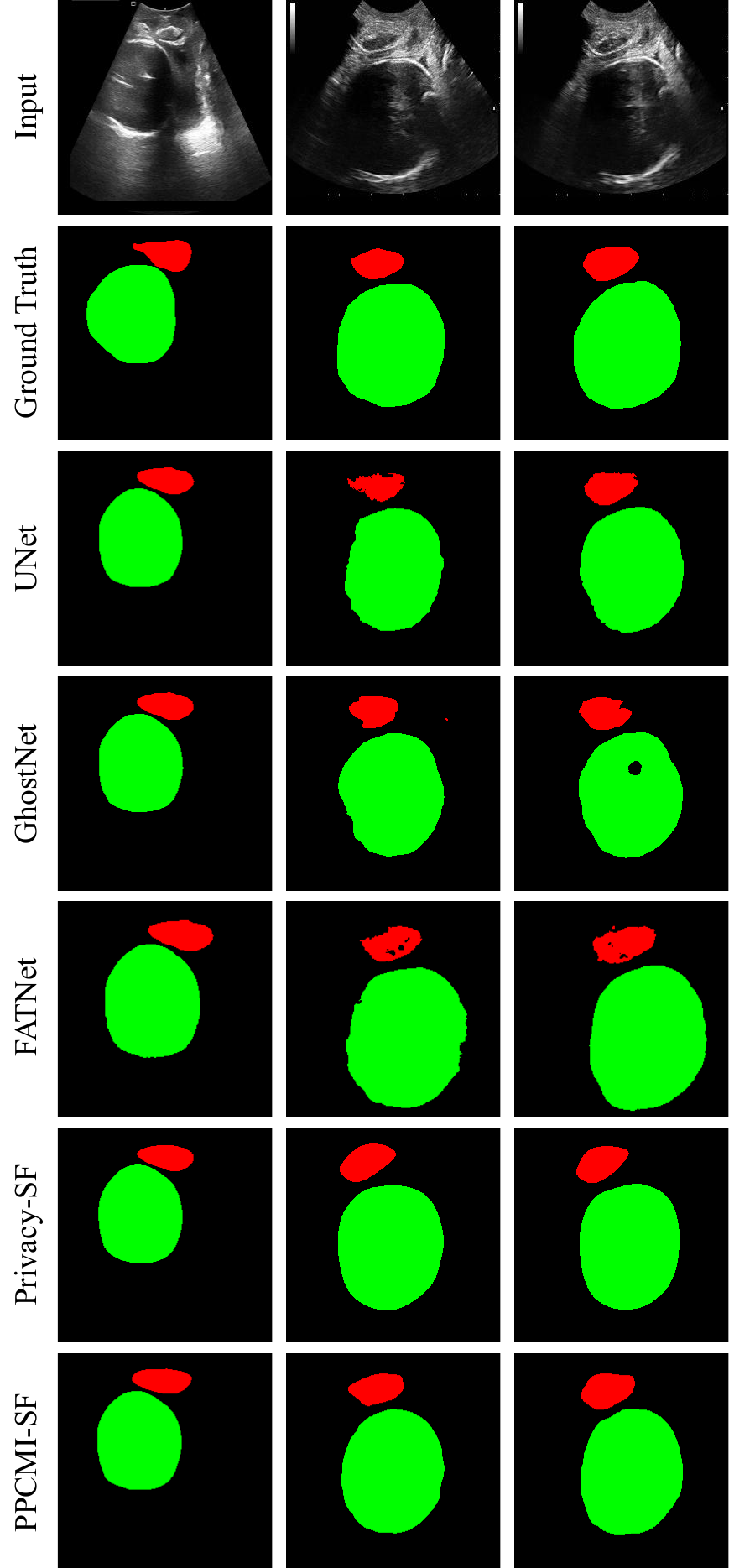}
\caption{Qualitative comparison of segmentation results on representative PSFH ultrasound images. Rows from top to bottom show the original input image, ground truth annotation, and predicted masks from UNet, GhostNet, FATNet, baseline Privacy-SF, and the proposed PPCMI-SF model, respectively. Fetal head and pubic symphysis regions are highlighted in green and red, respectively. }
\label{fig:psfh_visual}
\end{figure}

Table~\ref{tab:psfh_results} presents a comprehensive comparison of PPCMI-SF with state-of-the-art segmentation models on the PSFH dataset. Relative to the privacy-preserving baseline Privacy-SF, PPCMI-SF achieves consistent improvements across most evaluation metrics. The overall PSFH Dice score increases from $87.60 \pm 0.07$ to $90.49 \pm 0.05$, reflecting improved region overlap, while surface-based metrics also improve, indicating enhanced boundary localization and structural consistency. These gains demonstrate that the proposed architectural refinements effectively address the representation bottleneck observed in baseline Privacy-SF, enabling rich multi-scale feature preservation and more stable contour delineation within the protected latent domain.

When compared with privacy-agnostic models that operate directly in image space, PPCMI-SF remains highly competitive despite enforcing strict data isolation. Although several transformer-based approaches achieve slightly higher peak DSC values, they rely on full data accessibility and substantially larger model capacities. In contrast, PPCMI-SF attains comparable segmentation accuracy with a compact parameter footprint of only 20.65M, while maintaining latent-space protection throughout training and inference. The resulting HD95 and ASD values remain within the range of powerful privacy-agnostic CNN baselines, indicating that the proposed privacy-preserving design does not incur a significant penalty in boundary precision. Overall, these results highlight that PPCMI-SF narrows the performance gap between privacy-preserving and conventional segmentation frameworks, demonstrating that strong data protection and high segmentation fidelity can be achieved simultaneously.

The qualitative comparisons shown in Fig.~\ref{fig:psfh_visual} further support the quantitative results. UNet and GhostNet generally capture the overall fetal head morphology, however, boundary leakage is observed in low-contrast and acoustically shadowed regions. Privacy-SF exhibits more pronounced contour fragmentation around the pubic symphysis, consistent with the high HD95 and ASD values. In contrast, the proposed PPCMI-SF produces smoother and more anatomically consistent boundaries, with fewer spurious predictions and improved contour continuity. These visual improvements align with the improvements in the quantitative metrics and demonstrate that the proposed architectural enhancements preserve better spatial structure within the protected latent domain. Overall, the results indicate that the proposed PPCMI-SF effectively reduces the performance gap between privacy-preserving and privacy-agnostic models, achieving strong region overlap and accurate boundary delineation while operating under strict privacy constraints.

\subsection{Robustness and Cross-Dataset Generalization}
\label{sec:robustness_results}

\begin{figure}[h!]
\centering
\includegraphics[width=0.70\linewidth]{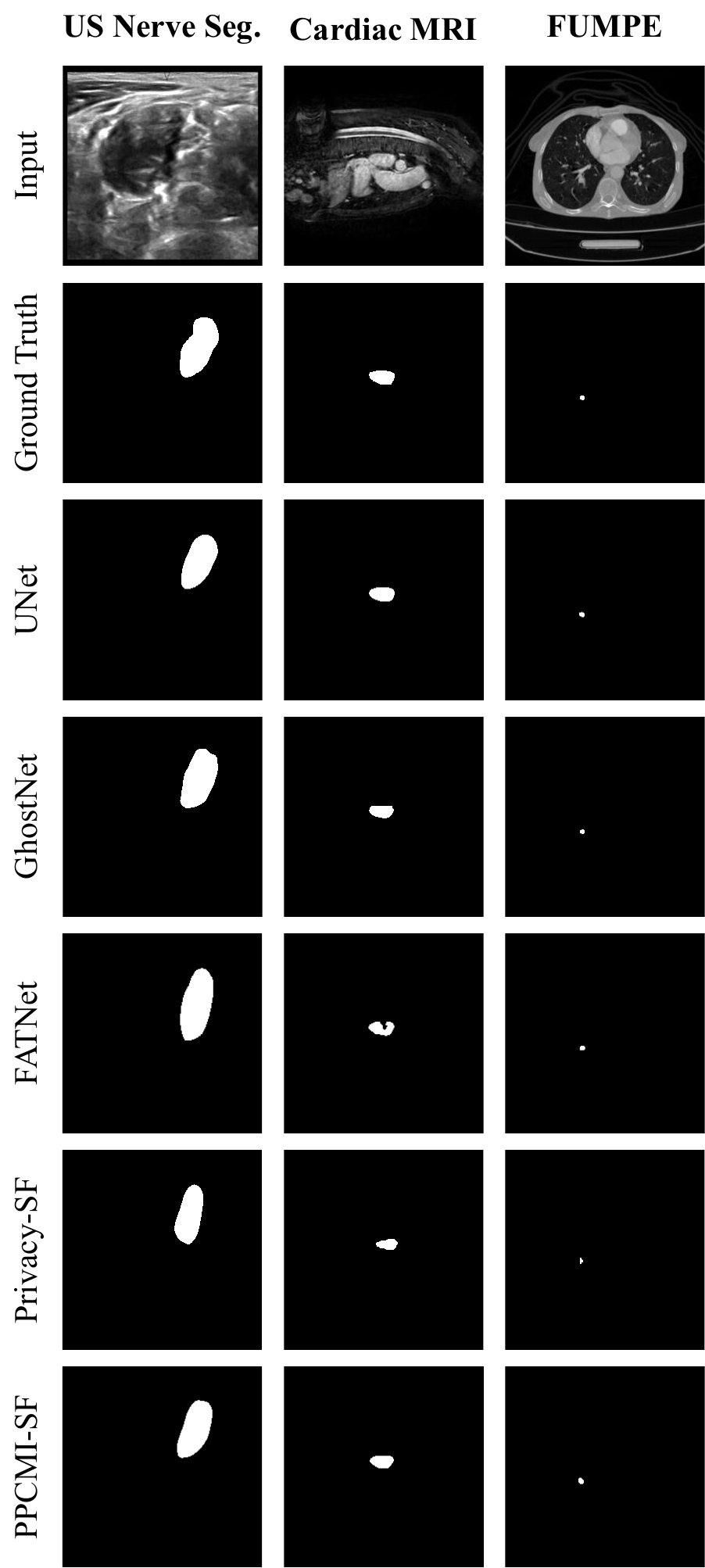}
\caption{Qualitative segmentation results across different datasets. The proposed PPCMI-SF maintains high structural fidelity while preserving privacy and closely tracks the performance of non-private UNet and GhostNet CNN baselines.
}
\label{fig:cross_dataset_visuals}
\end{figure}

\renewcommand{\arraystretch}{1.0}

\begin{table*}[!ht]
\centering
\caption{Quantitative comparison across datasets using dice similarity coefficient (DSC $\uparrow$), 95th percentile Hausdorff distance (HD95 $\downarrow$), and average symmetric surface distance (ASD $\downarrow$). Best results in each column are highlighted in bold.\\
}
\label{tab:cross_dataset_excel_formatted}
\resizebox{0.9\linewidth}{!}{
\begin{tabular}{c|ccc|ccc|ccc}
\hline \hline
\multirow{2}{*}{\bf Model} 
& \multicolumn{3}{c|}{\bf US Nerve Seg.} 
& \multicolumn{3}{c|}{\bf Cardiac MRI} 
& \multicolumn{3}{c}{\bf FUMPE} \\ \cline{2-10}

& \bf DSC $\uparrow$ & \bf HD95 $\downarrow$ & \bf ASD $\downarrow$
& \bf DSC $\uparrow$ & \bf HD95 $\downarrow$ & \bf ASD $\downarrow$
& \bf DSC $\uparrow$ & \bf HD95 $\downarrow$ & \bf ASD $\downarrow$ \\
\hline \hline

UNet \cite{ronneberger2015u}
& 80.32 & 12.01 & 4.14
& 88.40 & \textbf{2.06} & \textbf{0.63}
& 74.20 & 7.51 & \textbf{2.51} \\ \hline

GhostNet \cite{kazerouni2021ghost}
& 76.57 & 20.87 & 6.10
& 86.23 & 2.50 & 0.76
& 72.03 & 9.17 & 3.84 \\ \hline

FATNet ~\cite{wu2022fat}
& \textbf{80.55} & \textbf{10.56} & \textbf{3.93}
& 87.05 & 2.20 & 0.69
& 72.08 & 8.13 & 3.59 \\ \hline

Privacy-SF (Baseline) \cite{chen2024privacy}
& 69.45 & 13.13 & 5.79
& 78.94 & 4.98 & 2.13
& 61.34 & 11.41 & 6.41 \\ \hline

Proposed PPCMI-SF
& 76.90 & 11.23 & 4.25
& \textbf{88.87} & 3.47 & 1.18
& \textbf{74.54} & \textbf{6.30} & 3.90 \\ \hline \hline

\end{tabular}
}
\end{table*}

To evaluate the generalization performance beyond PSFH, all models were further evaluated on three additional datasets spanning distinct imaging modalities and anatomical targets: US Nerve Seg. dataset, FUMPE dataset, and Cardiac MRI dataset. These benchmarks introduce substantial domain shifts in appearance, noise characteristics, anatomical scale, and target morphology. All methods were trained and evaluated independently under identical experimental protocols, and the quantitative results are presented in Table~\ref{tab:cross_dataset_excel_formatted}.

On the US-Nerve Seg. dataset characterized by strong speckle noise, low contrast, and small target structures, the proposed PPCMI-SF achieves a DSC of $76.9\%$, improving over Privacy-SF ($69.4\%$) and comparable reductions in HD95 and ASD, indicating better boundary localization and reduced contour fragmentation. Although the privacy-agnostic baselines such as UNet and FATNet achieve higher DSC, the proposed PPCMI-SF significantly narrows the performance gap between privacy-preserving and privacy-agnostic models. The minor performance difference may be caused by the extreme sensitivity of small-structure segmentation to pixel-level details, which are partially attenuated by latent-space protection. Despite the performance difference, the reduced surface distance errors confirm that PPCMI-SF preserves boundary consistency more effectively compared to the baseline under challenging ultrasound conditions.

On the Cardiac MRI dataset, which requires precise delineation of ventricular cavities and myocardial boundaries under substantial inter-patient variability, the proposed PPCMI-SF achieves a DSC of $88.87\%$, slightly exceeding UNet and closely matching FATNet. Although UNet reports marginally lower boundary errors, the differences in HD95 and ASD remain small, indicating that the proposed method maintains competitive contour accuracy despite operating in the protected compressed latent domain. On the FUMPE CTA dataset, the propsoed PPCMI-SF attains the highest DSC of $74.54\%$ and the lowest HD95 of $6.30$, outperforming both the privacy-preserving baseline and conventional, privacy-agnostic CNN architectures. The improved boundary precision and overlap suggest effective localization of small embolic regions within complex vascular structures. These results demonstrate that the unified latent-space mapping combined with pyramid pooling preserves both local structural detail and global contextual information, enabling robust cross-modality generalization without compromising segmentation fidelity.

Fig.~\ref{fig:cross_dataset_visuals} presents qualitative performance evaluation consistent with the quantitative results. Across all evaluated datasets, PPCMI-SF generates anatomically coherent segmentations that closely align with the ground truth. Relative to Privacy-SF, the proposed method shows reduced fragmentation in nerve structures, fewer spurious artifacts in FUMPE CTA, and more clearly defined ventricular and myocardial boundaries in cardiac MRI. These observations indicate improved structural consistency under domain shifts.

\subsection{Effect of Multi-Client Collaboration}
\label{sec:multiclient_results}

\begin{figure}[h!]
\centering
\includegraphics[width=\linewidth]{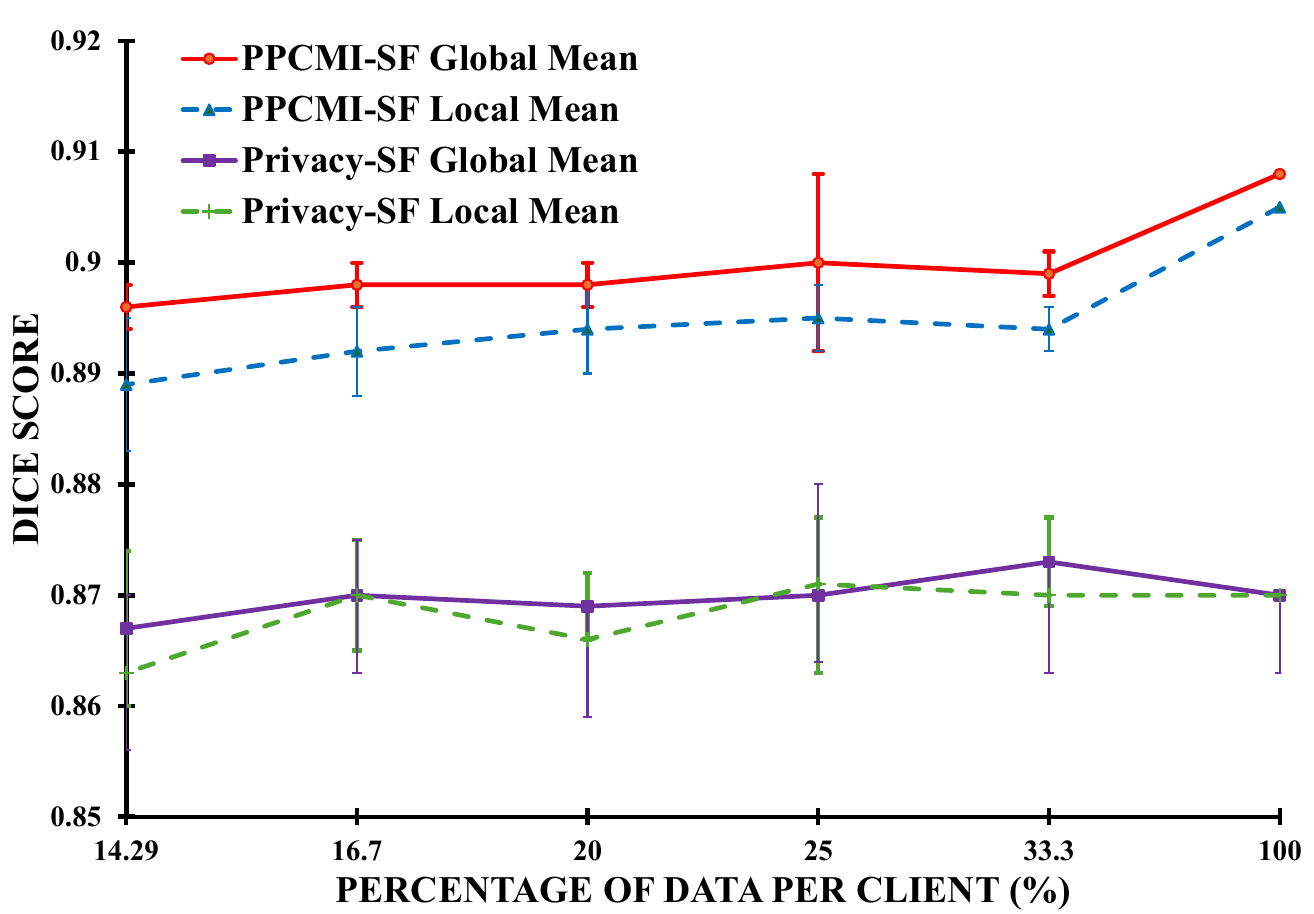}
\caption{Effect of per-client data reduction on segmentation Dice score under global collaborative training and local training on the PSFH dataset. The total training dataset size is fixed (4000 samples), while the number of participating clients increases, resulting in a decreasing percentage of data available per client. Error bars denote standard deviation across clients.}
\label{fig:multiclient_ablation}
\end{figure}

To evaluate scalability under realistic federated conditions, an ablation study is conducted to analyze the performance of the proposed method by varying the number of participating clients while keeping the size of the training set fixed at 4000 samples. As the number of clients increase, the per-client data fraction decrease, resulting in smaller local datasets for training each client-side autoencoder. Two collaboration strategies were considered: (1) \emph{global collaborative training}, where a single UMN is trained using aggregated latents from all clients, and (2) \emph{local training}, where each client trains an independent UMN using only local latents. Both the proposed PPCMI-SF and the baseline Privacy-SF were evaluated under identical settings on the PSFH dataset. Segmentation performance was measured using the DSC. 

As shown in Fig.~\ref{fig:multiclient_ablation}, the proposed PPCMI-SF maintains stable performance as per-client data decreases from 100\% to 14\%. Under global collaborative training, DSC decreases minimally from approximately 90.8\% to 89.6\%, indicating that the shared UMN effectively integrates heterogeneous latent representations into a unified manifold with strong generalization. In contrast, Privacy-SF exhibits consistently lower DSC and greater sensitivity to data reduction, with higher variance, particularly under global training. This behavior reflects the limited expressiveness of bottlenecked autoencoders without skip-connections, which struggle to preserve spatial context under data scarcity. For both methods, global training outperforms local training at all data fractions, with the gap widening as per-client data decreases, underscoring the benefit of collaborative learning. The global–local performance gap is larger for the proposed PPCMI-SF, indicating more effective exploitation of inter-client diversity. In the 3-client configuration ($n = 700$), paired DSC differences were non-normal (Shapiro–Wilk, $p < 0.001$), and the Wilcoxon signed-rank test showed that global training achieved a higher mean DSC (89.09\% vs. 88.74\%), with a mean improvement of 0.355 ($p = 0.0028$, $r = 0.113$). These results indicate that the proposed PPCMI-SF scales robustly as client numbers increase and per-client data decreases, sustaining segmentation quality under realistic multi-center conditions.

\subsection{Ablation Study of Key Architectural Components}
\label{sec:arch_ablation}

\begin{table*}[h!]
\centering
\caption{Ablation study results across PS, FH, and PSFH datasets using Dice (\%), HD95, and ASD metrics.}
\label{tab:ablation_architecture}
\renewcommand{\arraystretch}{1.15}
\setlength{\tabcolsep}{4.5pt}
\resizebox{0.8\linewidth}{!}{
\begin{tabular}{c|ccc|ccc|ccc|cc}
\hline \hline
\multirow{2}{*}{\bf Model} 
& \multicolumn{3}{c|}{\bf DSC (\%) $\uparrow$} 
& \multicolumn{3}{c|}{\bf HD95 $\downarrow$} 
& \multicolumn{3}{c|}{\bf ASD $\downarrow$}
& \multirow{2}{*}{\bf KLT} 
& \multirow{2}{*}{\bf Skips} \\ \cline{2-10}

& \bf PS & \bf FH & \bf PSFH
& \bf PS & \bf FH & \bf PSFH
& \bf PS & \bf FH & \bf PSFH
&  &  \\
\hline \hline

Full model (PPCMI-SF)
& 80.82 & 91.78 & 90.49
& 13.20 & 17.13 & 16.56
& 4.02 & 5.39 & 4.90
& \checkmark & \checkmark \\ \hline

No KLT
& 80.82 & 91.78 & 90.49
& 13.20 & 17.13 & 16.56
& 4.02 & 5.39 & 4.90
& -- & \checkmark \\ \hline

No skips (Plain AE)
& 75.28 & 88.95 & 87.47
& 11.35 & 18.15 & 17.80
& 4.84 & 6.40 & 5.87
& \checkmark & -- \\ \hline

Baseline (Privacy-SF)
& 74.52 & 87.89 & 86.51
& 17.66 & 18.81 & 18.50
& 5.80 & 6.79 & 6.13
& -- & -- \\ \hline \hline

\end{tabular}
}
\end{table*}

Ablation experiments were performed on the PSFH dataset by selectively disabling the KLT and skip-connections to quantify the contributions of each architectural component. Segmentation performance was evaluated using DSC, HD95, and ASD across PS, FH, and combined PSFH tasks, with results averaged over three clients and summarized in Table~\ref{tab:ablation_architecture}. Removing skip-connections results in consistent DSC degradation and increased boundary errors across all anatomical targets, indicating that multi-scale feature reuse is critical for preserving spatial detail in the latent space. The architecture variant without both KLT and skip-connections achieves the weakest performance overall, confirming that bottlenecked representations lack structural support and are insufficient for accurate and stable segmentation.

The complete PPCMI-SF model provides a balanced compromise between segmentation accuracy and privacy preservation. Diasbling KLT yields the same segmentation performance as the full model as shown in Table~\ref{tab:ablation_architecture}. This is because the server UMN consistently operates on the unprotected shared-domain latents of the images and masks, and as such it is unaffected by KLT. Evidently, KLT is applied only before transmission of the latents, and are inverted upon reception, resulting in identical inputs to the unified mapping network regardless of its use. Therefore, KLT does not influence segmentation accuracy but instead functions purely as a transmission-level privacy mechanism. 


\subsection{Cross-Decoder Prediction and Privacy Robustness}
\label{sec:cross-decoder-prediction}

To assess robustness against model inversion attacks, cross-decoder prediction experiments were performed by pairing the encoder from one client with a decoder from another client. If latent representations can be reconstructed by an external decoder, an adversary could potentially recover private information through decoder reuse or approximation. The worst-case pairing for each framework is reported, defined as the encoder–decoder combination yielding the highest reconstruction fidelity (largest SSIM and PSNR), corresponding to the strongest potential leakage scenario. For the baseline Privacy-SF, the worst-case pairing (Enc-2 $\rightarrow$ Dec-3) achieves $SSIM = 0.69$ and $PSNR = 18.91~dB$, indicating partial recovery of recognizable anatomical structures by an external decoder. In contrast, the worst-case pairing for the proposed PPCMI-SF (Enc-1 $\rightarrow$ Dec-2) results in substantially lower fidelity, with $SSIM = 0.34$ and $PSNR = 12.06~dB$, producing visually distorted and anatomically uninformative reconstructions.

\begin{figure}[h!]
\centering
\includegraphics[width=0.90\linewidth]{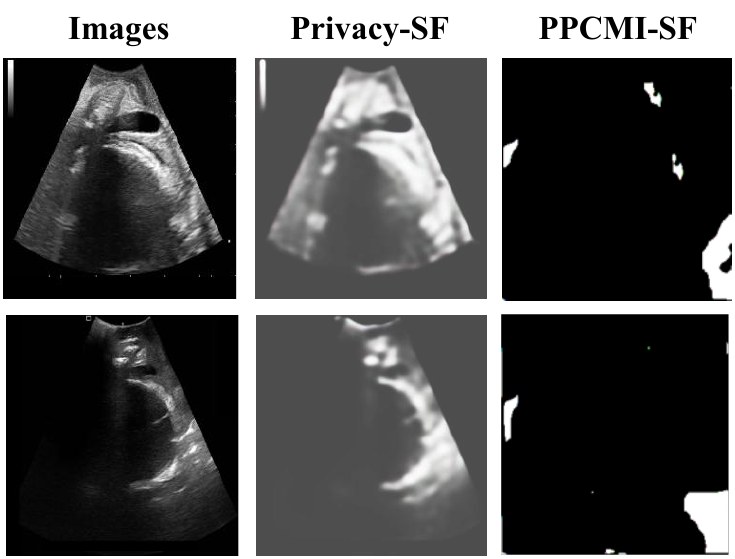}
\caption{Worst-case cross-decoder reconstructions for the proposed PPCMI-SF and baseline Privacy-SF on PSFH dataset. Privacy-SF exhibits partially recognizable anatomy under foreign decoder pairing, whereas PPCMI-SF produces visually distorted outputs, indicating stronger resistance to latent inversion.}
\label{fig:cross_pred_vis}
\end{figure}

Fig.~\ref{fig:cross_pred_vis} presents the visual results of the worst-case cross-decoder reconstructions. In the baseline Privacy-SF method, the reconstructed outputs retain identifiable anatomical structures, indicating that foreign decoders can partially recover meaningful content from shared latent representations. In contrast, the proposed PPCMI-SF produces heavily corrupted outputs that are visually disconnected from the original images. This behavior results from the client-specific KLT combined with enhanced latent regularization in the autoencoders, which prevents meaningful decoding by unintended decoders. These findings demonstrate that PPCMI-SF significantly mitigates the risk of model inversion through cross-client decoder reuse, thereby reinforcing privacy protection in collaborative environments.

\subsection{Membership-Inference Attack Evaluation}
\label{sec:membership_inference}

Membership-inference attacks (MIAs) attempt to determine whether a specific sample was included in a model training set, posing a significant privacy risk in medical imaging applications. Such attacks are particularly relevant in collaborative and federated learning settings, where model outputs or loss values may inadvertently encode information related to  individual training samples. To evaluate the robustness of the proposed PPCMI-SF against this threat, a loss-based MIA was performed targeting the server-side UMN across multiple architectural variants.

The adversary is modeled as a binary classifier that distinguishes member training samples from non-member held-out samples based on the prediction behavior of the mapping network. Performance is reported using the area under the ROC curve (AUC), sensitivity (true positive rate), specificity (true negative rate), and the \emph{Youden Index},
\[
(J = \text{Sensitivity} + \text{Specificity} - 1),
\]
where $\mathrm{AUC} \approx 0.5$ and $J \approx 0$ indicate performance close to random guessing, thereby reflecting strong resistance to membership inference.

\begin{table}[h!]
\centering
\caption{Membership-inference attack results for different architectural variants on the PSFH dataset. Lower AUC and Youden index indicate stronger privacy protection.}
\label{tab:mi_results}
\renewcommand{\arraystretch}{1.1}
\setlength{\tabcolsep}{5pt}
\begin{tabular}{l|c|c|c|c}
\hline \hline
\textbf{Variant} & \textbf{AUC} & \textbf{Sens.} & \textbf{Spec.} & \textbf{Youden $J$} \\
\hline \hline
PPCMI-SF (KLT + skips) 
& 0.473 & 0.548 & 0.413 & -0.039 \\
PPCMI-SF No KLT 
& 0.496 & 0.712 & 0.288 & -0.001 \\
PPCMI-SF No skips 
& 0.464 & 0.570 & 0.388 & -0.043 \\
Baseline (no KLT, no skips) 
& 0.470 & 0.610 & 0.372 & -0.018 \\
\hline \hline
\end{tabular}
\end{table}

As shown in Table~\ref{tab:mi_results}, all evaluated variants yield AUC values close to 0.5 and Youden indices near zero, indicating attack performance near random guessing. The complete PPCMI-SF model demonstrates strong resistance to membership inference, achieving an AUC of 0.473 and a Youden index of $-0.039$, suggesting that an adversary cannot reliably distinguish training samples from non-members at the selected decision threshold. In contrast, removing the KLT weakens membership privacy. The \emph{No KLT} variant attains the highest AUC of 0.496 and a near-zero Youden index, reflecting increased separability between member and non-member samples. Notably, sensitivity rises to $0.712$, indicating that training samples are detected substantially more frequently despite a reduction in specificity. This asymmetric behavior reveals measurable membership leakage. The \emph{No skips} and \emph{baseline} variants exhibit intermediate behavior, implying that architectural stabilization alone is insufficient to suppress membership signals when latent-space protection is reduced or absent. Overall, these results indicate that KLT is the primary factor mitigating membership inference risk. Although its removal may yield marginal gains in segmentation accuracy, it leads to increased exposure of membership information, whereas the full PPCMI-SF configuration maintains a more favorable balance between segmentation performance and resistance to inference attacks.

\subsection{Computation and Communication Cost}
\label{sec:cost}

The computational and communication overhead of PPCMI-SF during inference was analyzed using latency measurements obtained on an NVIDIA RTX A4500 GPU with batch size 1. During inference, each client transmits three protected latent tensors of size $(h_\ell, w_\ell, c)$ to the server and receives three predicted tensors of identical dimensions. The total number of transmitted values per query is given by
\[
\text{Payload} = \sum_{\ell} 2\,h_\ell w_\ell c,
\]
where the factor of 2 accounts for both upload and download. For latent spatial resolutions of $8 \times 8$, $16 \times 16$, and $32 \times 32$, with corresponding channel dimensions of 256, 128, and 64 in 32-bit floating-point precision, the resulting communication cost is approximately 0.88\,MB per query. This payload can be reduced by half through the use of 16-bit precision, demonstrating that the framework maintains low communication overhead while operating in the protected latent domain.

\begin{table}[h!]
\centering
\caption{Measured runtime and communication cost per query on an NVIDIA RTX A4500 (batch size=1, Float32 precision).}
\label{tab:runtime}
\renewcommand{\arraystretch}{1.15}
\setlength{\tabcolsep}{5pt}
\begin{tabular}{l|c|c}
\hline \hline
\textbf{Component} & \textbf{Time (ms)} & \textbf{Payload (MB)} \\
\hline \hline
Client encode \(E_x\) + KLT & 2.48 & 0.44  \\
Server mapping \(M\) & 14.03 & -- \\
Client inverse KLT + \(D_y\) & 2.56 & 0.44 \\
\hline
Total & 19.07 & 0.88 \\
\hline \hline
\end{tabular}
\end{table}

Table~\ref{tab:runtime} provides a breakdown of the end-to-end inference latency. Client-side encoding and KLT require approximately 2.48\,ms, server-side latent mapping takes 14.03\,ms, and client-side inverse KLT with decoding adds 2.56\,ms, resulting in a total latency of about 19.07\,ms per query. This runtime falls within the range of real-time or near-real-time clinical operation and is achieved without transmitting raw medical data. The proposed framework therefore combines strong segmentation accuracy with resistance to inversion and membership-inference attacks, while maintaining cross-dataset generalization and practical computational and communication efficiency, supporting its applicability in privacy-preserving multi-institutional medical imaging scenarios.

\section{CONCLUSION}
\label{sec:conclusion}

This study introduced PPCMI-SF, an enhanced privacy-preserving segmentation framework for secure multi-institutional medical imaging. By combining skip-connected autoencoders, a unified Pyramid Pooling--based mapping network, and a client-specific Keyed Latent Transform (KLT), the architecture protects latent representations while maintaining high segmentation fidelity. Extensive evaluations on ultrasound, CT, and MRI datasets demonstrate consistent accuracy improvements over the Privacy-SF baseline, robustness against cross-decoder reconstruction and membership-inference attacks (AUC $\approx$ 0.5), real-time inference (19\,ms latency), and low communication overhead ($<$1\,MB per query). These results indicate that strong empirical privacy and high utility can be jointly achieved through non-invertible latent transformations. While PPCMI-SF demonstrates strong empirical privacy and segmentation performance, it relies on assumptions about key confidentiality and server behavior. Future work will investigate integration with trusted execution environments, formal privacy mechanisms, and adaptive key rotation to further strengthen security guarantees in adversarial deployment settings.

\section*{Availability of Code}

The implementation of our proposed method is available on the following GitHub repository: \url{https://github.com/Shujaat123/Improved-Secure-SF}.

\section*{Acknowledgment}
Shujaat Khan acknowledges the support from the King Fahd University of Petroleum \& Minerals (KFUPM) under Early Career Research Grant no. EC241027. 

\section*{References}

\end{document}